\documentstyle[11pt]{article}
\begin{document}
\newcommand{\tr}{ \mbox{tr}}
\newcommand{\be}{\begin{equation}}
\newcommand{\ee}{\end{equation}}
\newcommand{\ba}{\begin{eqnarray}}
\newcommand{\ea}{\end{eqnarray}}

\begin{center}
{\large \bf  Asymptotic behavior in a model
with Yukawa interaction
from Schwinger-Dyson equations}\\
\vspace*{5 true mm}
{\bf V.E. Rochev\footnote{E-mail address: rochev@ihep.ru}}\\
{\it Institute for High Energy Physics, 142280 Protvino, Russia}
\end{center}
{\small{ {\bf Abstract.} A system of
Schwinger-Dyson equations  for pseudoscalar  four-dimensional Yukawa model
in the two-particle approximation
 is investigated. The simplest iterative solution of the system corresponds to the
mean-field approximation (or, equivalently, to the leading order of $1/N$-expansion)
and includes a non-physical Landau pole in deep-Euclidean region for the pseudoscalar
propagator $\Delta$. It is argued, however, that a full solution may be free from
non-physical singularities and has the self-consistent asymptotic behavior 
$p^2_e\Delta\simeq C\,\log^{-4/5}\frac{p^2_e}{M^2}$. An  approximate solution
  confirms the positivity of $C$ and the absence of
 Landau pole.
}\\


\section{Introduction}
A definition of asymptotic behavior at  large momenta
 for the strictly renormalized 
four-dimensional models of quantum field theory (QFT) up to now
is an unsolved problem. The unique exception is the theory 
with non-Abelian gauge interaction, which is asymptotically free
in the framework of renormalization-group-improved perturbation theory.
 A solution of the problem of
asymptotic behavior for other models requires  going out the framework 
of the coupling-constant
 perturbative expansion.
The first  attempt to define the asymptotic behavior in  QFT
 was made by Landau and coworkers in the 1950s.
This investigation
 was based on the approximate solution of Dyson equations
with the   summation of leading logarithms, and 
 the result for quantum electrodynamics was as follows: the photon
propagator included the non-physical singularity in Euclidean region of
momenta \cite{Landau}. Then similar singularities were indicated in the 
model with Yukawa interaction \cite{Abrikosov}. Such singularities in
the Euclidean
region  violate general principles of  QFT and
are a serious problem for these models.
Further development   has demonstrated
that these non-physical singularities arise practically
 inevitably in the framework of any known non-perturbative methods:
at the renormalization-group summation,
  in the frameworks of  $1/N$-expansion and mean-field expansion, etc.
\footnote{See reviews \cite{Callaway} for the historical survey and further
references.}

A widespread opinion is formulated as a triviality
of the quantum field models that is  not  asymptotically free
in the sense of the improved coupling-constant perturbative expansion.
There is a rigorous theorem \cite{Triviality} that the
four-dimensional scalar field theory with $\phi^4$ interaction on
the lattice does not  have an interacting continuum theory as its
limit for zero lattice spacing, i.e. the theory is trivial.
 However,  this argument  is not fully
conclusive due to an  uncertainty   of the continuous limit in
this model
\cite{Weinberg}.

In spite of the serious evidence for the triviality of lattice
scalar theory,  the situation with triviality today is as vague
as before, and 
 recent papers in this topic maintain incompatible statements.
While   mainstream works  confirm the triviality scenario,
  Suslov  in a series  of works(see \cite{Suslov} and references therein) 
argues the non-trivial behavior for $\phi^4$-theory and  quantum
electrodynamics  in the strong-coupling region. In any case, 
a study of the triviality problem requires a non-perturbative
tool. 

In this paper we investigate a new non-perturbative 
approximation for a model with
  Yukawa interaction
-- two-particle approximation, or 2PA. This approximation was proposed
for scalar fields in \cite{Rochev}, and it is 
 the first non-trivial step of a sequence of general
$n$-particle approximations, which tends to the exact infinite
system of Schwinger-Dyson equations (SDEs) at $n\rightarrow\infty$.

The structure of the paper is as follows: in  section 2,
the necessary notations and definitions are given; SDEs
 for the
generating functional of Green functions are introduced in the
formalism of a bilocal fermion source\footnote{A formalism
  of the bilocal source 
 was first
elaborated in  QFT  
by Dahmen and Jona-Lasinio \cite{Dahmen}. 
}. 
 We consider using of the bilocal source  as a convenient choice
of the functional variable. In particular, this variable is very convenient
for the construction of the mean-field (MF) expansion,
 which is presented in section 2. 
The existence of the Landau
pole in the boson propagator of the leading approximation of the 
MF expansion
 is also demonstrated in this section.

 In  section 3, a general
construction of the approximation scheme for the system 
of SDEs is given.
  The renormalization
of the system of equations is made, iterative solutions
 and some supplement
simplifications are discussed.

 In section 4, the asymptotic solution of the
system at large Euclidean momenta  is
presented and the asymptotic behavior of the propagators
 at large momenta
is discussed. The boson propagator in this model possesses 
self-consistent behavior.
  Conclusions are presented in section 5.

\section{ Schwinger-Dyson equations, mean-field approximation and Landau pole}

Consider the  theory of a Dirac fermion field $\psi$ interacting 
with a pseudoscalar boson field $\phi$ in a
four-dimensional (1+3) space  with the Lagrangian
\be
{\cal L}=\bar \psi(i\hat \partial -m)\psi-\frac{1}{2}\phi(M^2+\partial^2)\phi
+g \bar\psi\Gamma\psi \phi
\label{lagrangian} \ee
Here
$\Gamma=i\gamma^5$ and $\hat \partial=\gamma^\mu\partial_\mu$.
 The renormalizability in all orders of the coupling-constant
expansion requires to supplement Lagrangian 
(\ref{lagrangian}) with an additional
term $\lambda\phi^4$ which corresponds to the self-interaction of the boson field. 
This term ensures a renormalization of the boson-boson scattering amplitude.  
In this work we are not concerned with this amplitude,
 and the renormalization of approximations
 considered below does not  require  including the corresponding counter-term.
For this reason we do not include the quartic
 interaction into consideration. 
Therefore, we shall consider the restricted Yukawa model neglecting the
self-interaction of the boson field, and the obtained results should be treated
as the first step to the study of the asymptotic behavior in a realistic
model of the boson-fermion interaction. 
In other words,
we shall consider the case $\lambda=0$ only.
 Also we shall believe   $m>0$ and $M^2>0$
and, therefore, do not discuss in this work the  problem of
dynamical mass generation by Yukawa interaction (see \cite{Kondo}).   

 The generating functional of
Green functions  can be written as a functional integral
\be G=\int
D(\psi,\bar\psi, \phi)\exp
i\biggl[\int dx\, {\cal L} -\int dxdy\,\bar\psi(y)\eta(y,x)
\psi(x)+\int dx \,j(x)\phi(x)\biggr],
 \label{G} \ee 
where $j(x)$ is a single boson source and
$\eta(x,y)$ is a bilocal fermion source.

 The translational invariance of the functional 
integration measure leads to  the functional-differential
SDEs for   generating functional $G$. In terms of the  logarithm
 $Z=\frac{1}{i}\log G$ these equations are: 
\ba
\delta(x-y)+
(i\hat\partial_x-m)i\frac{\delta
Z}{\delta\eta(y,x)}+g\Gamma\bigg[\frac{\delta^2 Z
}{\delta\eta(y,x)\delta j(x)}+i\frac{\delta Z}{\delta\eta(y,x)}
\,\frac{\delta Z}{\delta j(x)}\bigg]=
\nonumber \\
=\int dx_1
\eta(x,x_1) i\frac{\delta
Z}{\delta\eta(y,x_1)},
\label{SDE1}
\ea 
\be
\frac{\delta Z}{\delta j(x)}=\int dx_1 
\bigg[ \Delta_c(x-x_1) j(x_1)
-g\Delta_c(x-x_1)\tr\Big(\Gamma\,
\frac{\delta Z}{\delta\eta(x_1, x_1)}\Big)\bigg].
\label{SDE2}
\ee 
Here $\Delta_c=(M^2+\partial^2)^{-1}$.
We define also the fermion propagator
\be
S(x-y)=
i\frac{\delta Z}{\delta\eta(y, x)}\,\bigg|_{\eta=j=0},
\label{S}
\ee
the boson propagator
\be
\Delta(x-y)=\frac{\delta^2Z}{\delta j(y)
\delta j(x)}\bigg|_{\eta=j=0},
\label{Delta} \ee
the two-particle (four-point) fermion function 
\be
Z_2\left(\begin{array}{cc}x&y\\x'&y'\end{array}\right)
=i\frac{\delta^2 Z}{\delta\eta(y',x')
\delta\eta(y, x)}\,\bigg|_{\eta=j=0}
\label{Z_2} \ee
and the three-point function
\be
G_3(z|x, y)=
i\frac{\delta^2Z}{\delta\eta(y, x)\delta j(z)}\bigg|_{\eta=j=0}.
\label{G_3} \ee
Differentiations of  SDE (\ref{SDE2}) over $\eta$ and $j$ 
give us  the SDE for the three-point function
\be
G_3(z|x, y)=-g\int dz_1 \Delta_c(z-z_1)\Gamma\,Z_2
\left(\begin{array}{cc}x&y\\z_1    &z_1    \end{array}\right)
\label{SDE_G_3} \ee
and the SDE for the boson propagator
\be
\Delta(x-y)=\Delta_c(x-y)+ig\int 
dy_1\tr\big[\Gamma\,G_3(x|y_1, y_1)\big]\,\Delta_c(y_1-y).
\label{SDE_Delta} \ee
Excluding with  the help of SDE (\ref{SDE2})
 a differentiation over $j$ in SDE (\ref{SDE1}),
we obtain at $j=0$ the 
SDE for the generating functional:
\ba
\delta(x-y)+
(i\hat\partial_x-m)i\frac{\delta
Z}{\delta\eta(y,x)}=\int dx_1 \bigg\{i\eta(x,x_1) \frac{\delta
Z}{\delta\eta(y,x_1)}+
\nonumber \\
+g^2\Delta_c(x-x_1)\Gamma\Big[
i\frac{\delta Z}{\delta\eta(y, x)}
\tr\Big(\Gamma\,\frac{\delta Z}{\delta \eta(x_1, x_1)}\Big)+
 \frac{\delta
}{\delta\eta(y,x)}\tr\Big(\Gamma\,\frac{\delta Z}
{\delta\eta(x_1, x_1)}\Big)
\Big]\bigg\}
\label{SDE_eta} \ea
which contains only 
the derivatives over the bilocal
source $\eta$.
Switching off the source $\eta$ in (\ref{SDE_eta}),
 we have the SDE for the fermion propagator
\be
(m-i\hat{\partial}_x)\,S(x-y)=
\delta(x-y)+
ig^2\int dx_1 \Delta_c(x-x_1)\Gamma Z_2
\left(\begin{array}{cc}x&y\\x_1&x_1\end{array}\right)
\Gamma
\label{SDE_S} \ee
A differentiation of (\ref{SDE_eta}) over $\eta$ gives us 
(with the source being switched off) 
the 
SDE for the
two-particle fermion function
\ba
(m-i\hat{\partial}_x)\,Z_2
\left(\begin{array}{cc}x&y\\x'&y'\end{array}\right)
+
\delta(x-y')\,S(x'-y)=
\nonumber \\
=ig^2\int dx_1 \Big\{(\Gamma\,S(x-y))\,\Delta_c(x-x_1) Z_2
\left(\begin{array}{cc}x_1&x_1\\x'&y'\end{array}\right)
\Gamma+
\Gamma\Delta_c(x-x_1) Z_3
\left(\begin{array}{cc}x&y\\x_1&x_1\\x'&y'\end{array}\right)
\Gamma\Big\}
\label{SDE_Z_2} \ea
Here $ Z_3=i\frac{\delta^3 Z}{\delta\eta^3}\big|_{\eta=0} $
is the three-particle (six-point) fermion function. 
The derivation of equation (\ref{SDE_Z_2}) implies that 
$\;\tr (\gamma^5S)=0$, i.e. we suppose  parity conservation.

To construct the MF expansion, 
 we consider as a leading approximation for equation
 (\ref{SDE_eta}) the 
equation \ba
\delta(x-y)+
(i\hat\partial_x-m)i\frac{\delta
Z^{MF}}{\delta\eta(y,x)}=
\nonumber \\
i\int dx_1\Big\{\eta(x,x_1) \frac{\delta
Z^{MF}}{\delta\eta(y,x_1)}+
g^2\Delta_c(x-x_1)\Gamma\,
\frac{\delta Z^{MF}}{\delta\eta(y, x)}
\tr\Big[\Gamma\,\frac{\delta Z^{MF}}{\delta 
\eta(x_1, x_1)}\Big]\Big\} \label{SDE_MF} \ea
We call (\ref{SDE_MF}) the MF approximation, since this 
equation gives us the same equations for propagators and
the two-particle funcion as the MF 
 expansion for the generating functional (see 
\cite{Rochev} and  references therein).  

The MF  fermion propagator is
\be
S=S^c,
\label{S_MF} \ee
where $S^c=(m-i\hat{\partial})^{-1}$. 

Equation (\ref{SDE_MF})  gives the equation for
 the two-particle function

\ba
Z_2\left(\begin{array}{cc}x&y\\x'&y'\end{array}\right)
+
S^c(x-y')\,S^c(x'-y)
=\nonumber 
\\
=ig^2\int dx_1 dx_2(S^c(x-x_1)\,\Gamma\, 
S^c(x_1-y))\,\Delta_c(x_1-x_2)
Z_2\left(\begin{array}{cc}x_2&x_2\\x'&y'\end{array}\right)
\Gamma
 \label{Z2eq_MF} 
\ea
whose solution is

 \ba
 Z_2\left(\begin{array}{cc}x&y\\x'&y'\end{array}\right)
=
-S^c(x-y')\,S^c(x'-y)+
\nonumber \\
+ \int dx_1 dx_2 (S^c(x-x_1)\,\Gamma\,S^c(x_1-y)) 
f_{MF}(x_1-x_2)
(S^c(x'-x_2)\,\Gamma\,S^c(x_2-y')),
\label{Z2_MF}
 \ea
 where in the momentum space
 \be
\frac{1}{f_{MF}(p^2)}=\frac{i}{g^2}
\Delta_c^{-1}(p^2)+L^c (p^2),
 \label{f_MF}
 \ee and 
\be
L^c (p^2)=\int \frac{d^4q}{(2\pi)^4}\,
\tr[S^c(p+q)\,\Gamma\,S^c(q)\,\Gamma]
  \label{L_MF}
\ee
is the single fermion loop.

Taking into account SDEs (\ref{SDE_G_3}) 
and (\ref{SDE_Delta}), we obtain the
MF boson propagator
\be
\Delta(p^2)=\frac{i}{g^2}\,f_{MF}(p^2).
\label{Delta_MF} \ee
The above formulae  contain
divergent integrals and should be renormalized.

The simplest method of  renormalization of the above equations is the 
direct application of a regularization procedure in the spirit of
Bogolyubov R-operation \cite{BogShir}.
The unrenormalized mass operator is
\be
\sigma(p^2)=\Delta^{-1}(p^2)-\Delta^{-1}_c(p^2)=-ig^2L^c (p^2).
\label{sigma_MF} 
\ee
Then the renormalized mass operator is defined as 
\be
\sigma_r(p^2)= {\mbox reg}\, \sigma(p^2)=-ig^2\, L^c_r (p^2),
\label{sigmar_MF}
\ee
where
\be
 L^c_r(p^2)=L^c(p^2)-L^c(0)-p^2(L^c)'(0)=
\frac{ip^2}{8\pi^2}\,
\int_0^1 dz \log[1-z(1-z)\frac{p^2}{m^2}].
\label{Lcr}
\ee 
is the renormalized fermion loop ( for the easement
 of the following calculations we 
choose the normalization point at zero
momenta).
As a result, the renormalized boson propagator $\Delta_r$  is
\be
\Delta^{-1}_r(p^2)=M^2-p^2-ig^2L^c_r(p^2)
\label{Deltar_MF}
\ee

We can also  renormalize the MF approximation by introducing 
 counter-terms in the Lagrangian.   
 In correspondence with the standard
recipe, we consider  (\ref{lagrangian}) as the  renormalized Lagrangian, 
where $\psi, \phi, m, M$ and $g$ are now the 
renormalized fields, masses and coupling,
and add  counter-terms
\be
\Delta{\cal L}=-\frac{z_\phi  -1}{2}\phi\,\partial^2\phi-
\frac{\delta M^2}{2}\phi^2
 \label{counterterms}
\ee
which absorb the divergences.

The full  Lagrangian ${\cal L}_b={\cal L}+\Delta{\cal L}$ 
can be written as
\be
{\cal L}_b={\cal L}+\Delta {\cal L}=\bar \psi_b(i\hat \partial -m_b)
\psi_b-\frac{1}{2}\phi_b(M_b^2+\partial^2)\phi_b
+g_b\bar\psi_b\Gamma\psi_b \phi_b
 \label{lagrB}
\ee
where
\be
\psi_b=\psi, \; \phi_b=\sqrt{z_\phi  }\phi, \;
g_b=\frac{g }{\sqrt{z_\phi  }}, \; m_b=m, \;
 M^2_b=\frac{ M^2+\delta M^2}{z_\phi  }.
\label{bare}
\ee
Then all the above calculations are reproduced with 
 Lagrangian (\ref{lagrB}), 
and the normalization conditions are imposed on the
 renormalized propagator $\Delta_r$.

The normalization conditions for the propagator 
$\Delta_r(p^2)=z^{-1}_\phi\Delta_b(p^2)$ are
\be
\Delta_r^{-1}(0)=M^2, \;\; 
\frac{d}{dp^2}\Delta_r^{-1}|_{p^2=0}=1 \label{normD}.
\ee
These conditions define the mass-renormalization 
counter-term $\delta M^2$ and the field-renormalization
constant $z_\phi$.  Then the renormalized
boson propagator is defined by equation (\ref{Deltar_MF}) as above.

As it follows from equation (\ref{Deltar_MF}), the
renormalized boson propagator $\Delta_r$ possesses a 
non-physical singularity
(Landau pole) in the Euclidean region $p^2<0$
 at the point $p^2_e\equiv-p^2=M_L^2$, where $M^2_L$ is a
solution of the equation
$$
M^2+M_L^2-ig^2L^c_r(-M^2_L)=0. \label{Landaupole}
$$
This equation has a solution at any positive $g^2$.
 As  was yet
noted in the introduction, the  same Landau pole arises in the
calculations of the renormalized amplitude by other methods: in the
frameworks of  $1/N$-expansion and   renormalization-group
summation.

\section{The system of  SDEs and two-particle approximation}

 The system of fermion SDEs is an infinite set
of equations for $n$-particle fermion functions 
$Z_n\equiv i\,\delta^n
Z/\delta\eta^n|_{\eta=0}$. The first SDE is 
equation (\ref{SDE_S}). The second SDE is equation
 (\ref{SDE_Z_2}). The $n$th SDE
is the $(n-1)$th derivative of SDE (\ref{SDE_eta}) 
with the source
being switched off and includes a set of functions from
one-particle fermion function $S$ to $(n+1)$-particle 
fermion function
$Z_{n+1}$.

 We call "the $n$-particle approximation of the system of SDEs"
the system of  $n$ SDEs in which the first $n-1$
 equations are exact and the $n$th
SDE is truncated by omitting the $(n+1)$-particle function.
 It is evident that
the sequence of such approximations goes to the 
exact set of SDEs at $n\rightarrow\infty$.
The one-particle approximation is simply equation 
(\ref{SDE_S}) without $Z_2$. This approximation
has a trivial solution which is a free propagator. 
The two-particle approximation is a system of equation
(\ref{SDE_S}) and equation (\ref{SDE_Z_2}) without $Z_3$:
\ba
(m-i\hat{\partial}_x)\,Z_2
\left(\begin{array}{cc}x&y\\x'&y'\end{array}\right)
+
\delta(x-y')\,S(x'-y)=
\nonumber \\
=ig^2\int dx_1 (\Gamma\,S(x-y))\,\Delta_c(x-x_1) Z_2
\left(\begin{array}{cc}x_1&x_1\\x'&y'\end{array}\right)
\Gamma
\label{SDE_Z_2P} \ea
which includes $S$ and two-particle function $Z_2$. 
This nonlinear system will be
the  object of the present investigation.

The idea of these approximation scheme is very simple and
natural. However, the calculations became more and more
complicated at each following stage: e.g., the three-particle
approximation is a system of three nonlinear equations for the
propagator, the two-particle function and the three-particle
function. 

Another view to the origin of system (\ref{SDE_S}) and
(\ref{SDE_Z_2P}) is based on a modification of  the MF
expansion of  section 2
 with taking into account a particular
solution of functional-derivative equation (\ref{SDE_eta}). It is
easy to see that the SDE (\ref{SDE_eta})   has the simple 
solution \be
Z_{p}(\eta)=\frac{1}{2i}\,Z_2\cdot\eta^2-iS\cdot\eta,
\label{Zpart} \ee 
where functions $Z_2\left(\begin{array}{cc}
x&y\\x'&y'\end{array} \right)$ and $S(x-y)$ satisfy  the
system of equations (\ref{SDE_S}) and (\ref{SDE_Z_2P}).
To be an exact solution of the functional-derivative equation 
(\ref{SDE_eta}),
   this system should be supplemented by one more nonlinear
equation for $Z_2$. The system of three equations
 for the two functions
$S$ and $Z_2$ are overfull and seemingly has not  physically
meaningful solutions. However, the third equation
 does not play a role for the construction of the modified
MF expansion. The  construction of this expansion for 
scalar field theory can be found in  work \cite{Rochev}.
This construction  can be generalized also to  fermion fields.  
 At the $n$th step of this expansion,
we have a closed system of linear integral equations, and
therefore this scheme is much less complicated in the
 calculational sense
in comparison to
the above scheme of the $n$-particle approximations.
 Equations (\ref{SDE_S}) and
(\ref{SDE_Z_2P}) are the basic approximation for this expansion.

Equations (\ref{SDE_S}) and (\ref{SDE_Z_2P}) are 
the system of nonlinear equations
for the functions $S$ and $Z_2$. In equation (\ref{SDE_Z_2P}),
the  two-particle function $Z_2$ can be considered 
as a functional of $S$, and the "solution"
of this equation can be easily found:
\ba
Z_2\left(\begin{array}{cc}x&y\\x'&y'\end{array}\right)
=
-S^c(x-y')\,S(x'-y)
\nonumber \\
+\int dx_1 dx_2\,(S^c(x-x_1)\Gamma\,S(x_1-y)) f(x_1-x_2)
(S(x'-x_2)\Gamma\,S^c(x_2-y'))   \label{Z2}
\ea
Here
\be
f(x-y)=-ig^2\,\Delta_c(x-y)+ig^2\int dx_1 dx_2 \,
\Delta_c(x-x_1) L(x_1-x_2) f(x_2-y)
\label{f}
\ee
and $L (x)=\tr[S^c(x)\,\Gamma\,S(-x)\,\Gamma]$
is the fermion loop operator.

In  momentum space:
\be
\frac{1}{f(p^2)}=\frac{i}{g^2}\,\Delta^{-1}_c(p^2)+L (p^2),
\label{f_2P}
\ee
\be
L (p^2)=\int \frac{d^4q}{(2\pi)^4}\,
\tr[S^c(p+q)\,\Gamma\,S(q)\,\Gamma].
\label{L}
\ee
Taking into account equations 
(\ref{Z2})--(\ref{L}), (\ref{SDE_G_3}) 
and (\ref{SDE_Delta}),
 we obtain for the
boson propagator $\Delta$
the following equation 
in  momentum space:
\be
\Delta^{-1}(p^2)=M^2-p^2-ig^2L(p^2).
\label{Delta_2P}
\ee
From equation (\ref{SDE_S}) and equations 
(\ref{Z2})--(\ref{Delta_2P})
 we have the  equation for the fermion propagator 
\be
S^{-1}(p)=m-\hat{p}+ig^2\,K(p),
\label{S_2P}
\ee
where
\be
K(p)=\int \frac{d^4q}{(2\pi)^4} \,\Gamma\,S^c(p-q)\,\Gamma\Delta(q).
\label{K}
\ee
The system of equations (\ref{L})--(\ref{K}) is the system of 
unrenormalized SDEs in the 2PA.

The renormalization of  equations (\ref{Delta_2P}) and (\ref{S_2P})
 can be performed in correspondence with the
general recipe of section 2.

If we define the 
unrenormalized  mass operators as
\be
\cases{\Sigma(p)=S^{-1}(p)-S^{-1}_c(p)=ig^2K(p)
\cr \sigma(p^2)=\Delta^{-1}(p^2)-\Delta^{-1}_c(p^2)=-ig^2L(p^2)}
\label{MO}
\ee
then the renormalized  mass operators are
\be
\cases{\Sigma_r(p)\equiv \mbox{reg} \;\Sigma(p)=
ig^2\bigg(K(p)-K(0)-\hat{p}\,
\frac{\partial K}{\partial\hat{p}}\bigg|_{p=0}\bigg)
\cr \sigma_r(p^2)\equiv \mbox{reg} \; \sigma(p^2)
=-ig^2\Big(L(p^2)-L(0)-p^2L'(0)\Big)},
\label{MOr}
\ee
and the system of renormalized equations for propagators are
\be
\cases{S^{-1}_r(p)=m-\hat{p}+ig^2\,K_r(p)
\cr
 \Delta^{-1}_r(p^2)= M^2-p^2-ig^2\,L_{r}(p^2) }
\label{2PS}
\ee
where
\be
K_r(p)=K(p)-K(0)-\hat{p}\frac{\partial K}
{\partial\hat{p}}\bigg|_{p=0},
\label{Kr}
\ee
\be
L_{r}(p^2)=L(p^2)-L(0)-p^2L'(0).
\label{Lr}
\ee

 The renormalization with counter-terms can be performed by introducing 
the counter-term Lagrangian 
\be
\Delta {\cal L}=(z_\psi-1)\bar \psi\, i\hat \partial\psi -\delta m\,
\bar{\psi}\psi-\frac{z_\phi  -1}{2}\phi\,\partial^2\phi-
\frac{\delta M^2}{2}\phi^2
\label{counter_2P}
\ee
and full Lagrangian (\ref{lagrB}).
The bare quantities in (\ref{lagrB}) are now
\be
\psi_b=\sqrt{z_\psi}\psi, \; \phi_b=\sqrt{z_\phi  }\phi,
\;
g_b=\frac{g }{z_\psi\sqrt{z_\phi  }},
 \; m_b=\frac{m+\delta m}{z_\psi}, \;
 M^2_b=\frac{ M^2+\delta M^2}{z_\phi}.
\ee
The bare mass operators $\Sigma_b$ and $\sigma_b$ in the
 2PA are given
by formulae (\ref{MO}) with substitutions $\Sigma\rightarrow\Sigma_b,\, 
\sigma\rightarrow\sigma_b,\, g\rightarrow g_b,\, m\rightarrow m_b, \, M^2\rightarrow M^2_b$,
etc.
Normalization conditions (\ref{normD}) 
 for the propagator $\Delta_r(p^2)$ 
 and for
the fermion propagator 
\be
S^{-1}_r(p=0)=m, \;\; 
\frac{\partial S^{-1}_r(p)}{\partial\hat{p}}\bigg|_{p=0}=-1
\label{normS}
\ee
 define the  counter-terms,
  and the system of renormalized
equations will be system (\ref{2PS}) again.

Note that  an iteration of equation for $\Delta_r$ in
 (\ref{2PS}) with $S^{(0)}=S^c$
leads to the MF propagator (\ref{Deltar_MF}).
 So the MF approximation and the
equivalent leading-order $1/N$--expansion are contained in the
2PA as the first iteration.

Equations (\ref{2PS}) are the system of nonlinear integral
equations for the propagators.
The most interesting problem is to look for the 
asymptotic behavior of the solution 
of system (\ref{2PS}) at large Euclidean
momenta.
In the large-momenta region, an essential technical  simplification
 is possible, namely, one can
replace in integrals (\ref{L}) and (\ref{K}) the function $S_c$
 by a massless function $-1/\hat{p}$:
\be
\int \frac{d^4q}{(2\pi)^4} F(q)\,S_c(p-q)
\Longrightarrow-\int \frac{d^4q}{(2\pi)^4}
\,\frac{ F(q)}{\hat{p}-\hat{q}}.
\label{zeromass}
\ee
Then it is possible to use the well-known formula
\be
\int \frac{d^4q_e}{(2\pi)^4}
\,\frac{ f(q^2_e)}{(p-q)^2_e}=\frac{1}{16\pi^2}\bigg[\frac{1}{p^2_e}
\int_0^{p^2_e} f(q^2_e)\, q^2_e\, dq^2_e+
\int_{p^2_e}^\infty f(q^2_e)\,dq^2_e\bigg].
\label{Int}
\ee
This  massless-integration approximation (\ref{zeromass})
 is quite usual in investigations in the deep-Euclidean region,
though rigorous arguments for its validity can be done for
the asymptotically-free models only \cite{Weinberg}.
In the general case, this approximation should be considered
as a plausible conjecture, which needs  further investigations.
Formula (\ref{Int}) highly enables the calculations and,
 as a major point, permits us to go
from integral equations to differential ones (see below).

 Equations (\ref{2PS}) in the massless-integration approximation in
the Euclidean region are
\be
\cases{ a(p^2_e)=1-\frac{g^2}{32\pi^2(p^2_e)^2}
\int_0^{p^2_e} dq^2_e(p^2_e-q^2_e)^2\,\Delta_r(q^2_e)
\cr \Delta^{-1}_r(p^2_e)= M^2+p^2_e-\frac{g^2}{8\pi^2 p^2_e}
\int_0^{p^2_e} dq^2_e(p^2_e-q^2_e)^2\,\frac{a(q^2_e)}{m^2+q^2_ea^2}
}
\label{2PSe}
\ee
Here $a$ is defined by $S^{-1}_r$ with formula 
$$
S^{-1}_r=b-a\hat{p},
$$
and $b=m$ in the massless-integration approximation.
Euclidean normalization conditions are
\be
\Delta^{-1}_r(0)=M^2, \; (\Delta^{-1}_r)'(0)=1, \; a(0)=1. 
\label{EuNorm}
\ee
Introducing the dimensionless quantities
$$
t=\frac{p^2_e}{M^2},  \; \mu^2=\frac{m^2}{M^2}, \;
\; h(t)=\frac{1}{p^2_e\Delta_r},
$$
 system (\ref{2PSe}) can be written as follows
\be
\cases{a(t)=1-\frac{g^2}{32\pi^2}\,\int_0^t\,(1-\frac{t_1}{t})^2\,
 \frac{dt_1}{t_1h(t_1)}
\cr
h(t)=1+\frac{1}{t}-\frac{g^2}{8\pi^2}\int_0^t\, (1-\frac{t_1}{t})^2\,
\frac{dt_1\,a(t_1)}{\mu^2+t_1a^2(t_1)} }
\label{IntSystem} \ee

Let us discuss the iterative solutions of system (\ref{IntSystem}).
 Due to the nonlinearity of equations an essential moment is a way of 
iterations. Consider two schemes  of iterations. The first scheme is a choice
$a^{(0)}=1$ as a leading order, and then the calculation of $h^{(1)}$ from
the second equation with $a=a^{(0)}$. As  was pointed above,
 this scheme corresponds to the MF
approximation  and leads to the Landau pole in the boson propagator $\Delta_r$ at
 point $t_L$.  At this point $h^{(1)}(t_L)=0$, 
and the following
calculation of $a^{(1)}$ from the first equation is possible only
at $t<t_L$. For this reason the subsequent calculations in the
framework of this scheme became  problematical  in the region of large $t$.

 Another iterative scheme is a choice of the leading order as $h_0=1+\frac{1}{t}$,
and the following calculation of $a_1$ from the first equation with $h=h_0$:
\be
a_1=-\frac{g^2}{32\pi^2}\Big(1+\frac{1}{t}\Big)^2\log(1+t)+
1+\frac{3g^2}{64\pi^2}+\frac{g^2}{32\pi^2t}.
\label{a1}
\ee
Likewise $h^{(1)}$ in the first scheme,  function $a_1(t)$  equals to zero 
at some  point $t_0$, but consequences of this fact are
quite different. Since $\mu^2+ta^2_1>0$ at any $t>0$,\footnote{A case
$\mu^2=0$ needs  a special consideration and is not discussed here.} then $h_1$
can be calculated for any $t$, where at $t\rightarrow\infty$
 \be
h_1\simeq  4\log\log t >0,
\label{h1}
\ee
i.e., $h_1$ has a self-consistent asymptotic behavior. Certainly, a finite number
of iteration cannot define the actual asymptotic behavior, but
the given considerations demonstrate the obvious preference of the second
scheme in comparison with
the usual MF approximation.
This preference consists in the absence of the Landau pole in the fermion
propagator, which is defined by $a_1$ of equation (\ref{a1}) at $\mu^2>0$.
This circumstance is essentially used for the construction of an approximate
solution in  the following section.

\section{Asymptotic behavior}
 System (\ref{IntSystem}) (multiplied to $t^2$) after threefold differentiations
  is reduced  to the
system of differential equations
\be
\cases{
\frac{d^3}{dt^3}(t^2a)=-\frac{g^2}{16\pi^2}\frac{1}{th}
\cr
 \frac{d^3}{dt^3}(t^2h)=-\frac{g^2}{4\pi^2}\frac{a}{\mu^2+a^2t} }
\label{DifSystem}
\ee
At $t\rightarrow 0$ system of
integral equations (\ref{IntSystem}) gives us:
\be
a(t) = 1-\frac{g^2}{96\pi^2}\,t+\frac{g^2}{384\pi^2}\,t^2+O(t^3)
\label{zero}
\ee
and
\be
h(t)= \frac{1}{t}+1-\frac{g^2}{24\pi^2\mu^2}\,t+O(t^2)
\label{hzero}
\ee
These formulae give us boundary conditions for system (\ref{DifSystem})
at the point $t=0$.

 At large $t$, system (\ref{DifSystem}) has  
the asymptotic solution\footnote{Note that   system (\ref{DifSystem})
 has  exact solution 
$a=C\,t^{-1/2}, \; h=\frac{g^2\,t^{1/2}}{6\pi^2C}, \; C^2=-\frac{5}{9}\mu^2.$ 
This  imaginary solution is not, of course, a solution  of
integral equations (\ref{IntSystem}).} :
\be
a\simeq {\cal A}\,\log^{1/5}t, \; h\simeq {\cal B}\,\log^{4/5}t, 
\;{\cal AB}=-\frac{5g^2}{32\pi^2}.
\label{AsSol}
\ee
Differential equations (\ref{DifSystem}) do not fix 
the signs of coefficients ${\cal A}$ and ${\cal B}$. These signs have
the principal 
 meaning and define the physical situation,  described 
by 2PA. If ${\cal A}>0 \, ({\cal B}<0)$, then function
$h(t)$ applies to zero at some point. This case corresponds to the
presence of  Landau-type singularity in the boson propagator, i.e., the
situation is similar to the above   physically
unsatisfactory MF approximation. 
Oppositely, if ${\cal A}<0 \, ({\cal B}>0)$, then the situation
 corresponds to  self-consistent asymptotic behavior of the boson propagator. 
In this case, function $a(t)$ has a zero at some point, but it does not lead to
a Landau pole in the propagator for  massive fermions. 

System (\ref{DifSystem}) of nonlinear differential equations is 
rather difficult
for  detailed analytical investigation, and below we make
 some approximations and
natural simplified suppositions, which will enable 
the construction of an approximate
solution and the definition of the coefficients  ${\cal A}$ and ${\cal B}$.

Firstly note, that  function $v=ah$ has  quite definite asymptotic 
behavior at $t\rightarrow\infty$:
\be
v=ah \simeq -\frac{5g^2}{32\pi^2}\,\log t.
\label{v}
\ee 
Since $v\simeq \frac{1}{t}$ at   $t\rightarrow 0$, this 
function changes the
sign in some point $t_0$ and $v(t_0)=0$.
 We shall suppose the uniqueness of this point. Consider the
behavior of the solution near the point $t_0$.    
After the change of  variable
$
x=\log t,
$
the first equation of system (\ref{DifSystem}) can be written as
\be
a'''+3a''+2a'=-\frac{g^2}{16\pi^2}\frac{1}{h}=
-\frac{g^2}{16\pi^2}\frac{a}{v}.
\label{DifSystem_x}
\ee
Here $a'\equiv \frac{da}{dx}$. Suppose in a vicinity of zero
 point $x_0=\log t_0$:
\be
v\approx v'(x_0)(x-x_0).
\label{vx0}
\ee
Note that supposition (\ref{vx0}) is fulfilled for iterative
solution (\ref{a1}), which will be used in the construction of the
approximate solution (see below).

Then, going to the variable $z=x-x_0=\log\frac{t}{t_0}, \;\; $ 
 we obtain for $a$ the linear 
differential equation 
\be
a'''+3a''+2a'= k  \,\frac{a}{z},
\label{a_z}
\ee
where
$ k  =-\frac{g^2}{16\pi^2}\frac{1}{v'(x_0)}.$
Since $v$ decreases, $ k  >0$.
With the substitution
\be
a=e^{-z}u
\label{au}
\ee
and by changing the variable
$\xi=\frac{z^2}{4}$,
we obtain for $u$ the Meijer equation \cite{Luke}
$$
[\xi(\xi\frac{d}{d\xi}
+\frac{ k  }{2})-(\xi\frac{d}{d\xi}-1)(\xi\frac{d}{d\xi}
-\frac{1}{2})\xi\frac{d}{d\xi} ]\,u=0,
$$
whose solution is
\be
u=C_1u_1+C_2u_2+C_3u_3,
\label{Meijer}
\ee
where
\ba
u_1=G^{11}_{13}\Big(-\frac{z^2}{4}\Big|
 \begin{array}{ccc} 1-k/2\\0; \frac{1}{2}, 1 \end{array}\Big),\;
u_2=i\,G^{11}_{13}\Big(-\frac{z^2}{4}\Big|
 \begin{array}{ccc} 1-k/2\\ \frac{1}{2}; 0, 1 \end{array}\Big),\;
\nonumber \\
u_3= G^{21}_{13}\Big(\,\frac{z^2}{4}\Big|
 \begin{array}{ccc} 1-k/2\\0, 1; \frac{1}{2} \end{array}\Big).
\label{u123}
\ea 
At $z\rightarrow 0$ we have
$u_1
\sim z^2, \; u_2
 \sim z, \; u_3\sim z^0.
$
Consequently, if $C_3\neq 0,$  function $a$ does not change 
sign at  point $x_0$, and ${\cal A}>0$. If $C_3=0, \, C_2\neq 0,$ 
function $a$ changes  sign, and ${\cal A}< 0$. This case corresponds
to the above-mentioned situation of  self-consistent asymptotic behavior.
The case $C_3=C_2=0$ corresponds to  touching
 for $a$ and a pole singularity
for $h$.

A definition of coefficients $C_i$ needs  some boundary conditions.
These boundary conditions should be connected with boundary
 conditions at $t=0$ and 
can be defined on the basis of some
approximate solution in the  region of small $t$. The linearized
version type of equation (\ref{a_z}) is tightly connected
 with the asymptotic behavior
at large $t$ and  apparently cannot be applied in the region of small $t$.
For this reason, in  the pre-asymptotic region we shall 
use iterative solution
(\ref{a1}), which satisfies 
supposition (\ref{vx0}) and  boundary conditions
 (\ref{zero}). In other words, we believe
$
a= a_1
$
at $t\leq t_0 \; (x\leq x_0)$.
In the region of large $t,$ we  use asymptotic formula (\ref{v}) for $v$
and believe 
$
v= -\frac{5g^2}{32\pi^2}(x-x_0)
$
at $t\geq t_0 \;(x\geq x_0)$.
Correspondingly,  we have for $a$ at large $t$ equation (\ref{a_z}),
 whose solution is
 given by equations (\ref{au})--(\ref{u123})  with $ k  =2/5$.

As the boundary conditions we shall use the conditions
 of ``a smooth join''
at  point $x=x_0 \; (z=0)$:
\be
 a(z=0)=a(x_0)=a_1(x_0)=0, \; \; a'_z(0)=a'_1(x_0), 
\; \; a''_{zz}(0)=a''_1(x_0).
\label{join}
\ee 
From these boundary conditions, we define 
\be
 C_1=\frac{\sqrt{\pi}}{\Gamma(1.2)}\,(a''_1(x_0)+2a'_1(x_0)),\;\;
  C_2=\frac{\pi}{\Gamma(0.7)}\,a'_1(x_0),\;\;C_3=0.
\label{C}
\ee
Here $\Gamma(\kappa)$ is the Euler gamma-function.
Simple calculation shows that both $C_1$ and $C_2$ are negative.
Asymptotic expansions of Meijer functions (see \cite{Luke}) give
us the behavior of solution 
at $z\rightarrow+\infty$:
\be
u=C_1u_1+C_2u_2\simeq e^z\,\Big(\frac{z}{2}\Big)^{1/5}\,
\frac{1}{2\sqrt{\pi}}\,(C_2+C_1),
\label{uAs}
\ee
and, consequently, the asymptotic behavior of this 
approximate solution is
given by formulae (\ref{AsSol}) with 
\be
{\cal A}=\frac{1}{2^{6/5}\sqrt{\pi}}\,(C_1+C_2)<0
\label{A}
\ee
and ${\cal B}>0$. 
At $t_0\gg 1$
\be
{\cal A}
\approx -0.77\,\frac{g^2}{16\pi^2},\; \;{\cal B}\approx 3.24,
\label{AB}
\ee
and
\be
\Delta_r\simeq\frac{0.308}{p^2_e\log^{4/5}\frac{p^2_e}{M^2}}.
\label{DeltaAs}
\ee

\section{Conclusions}
Our results  demonstrate that two-particle
approximation of the system of Schwinger-Dyson equations essentially differs
in the asymptotic deep-Euclidean region of momenta in comparison with the MF approximation.
Instead of physically unsatisfactory behavior with the Landau pole in the
Euclidean region, which occurs for the MF approximation, or for the
leading term of $1/N$-expansion, the boson propagator in the two-particle
approximation has self-consistent asymptotic behavior, which is similar to
the asymptotically free behavior. Certainly,  inclusion of the scalar 
self-action into consideration can notably vary the results in the
asymptotic region. From this point of view, the obtained results should
be considered as the first step of investigation of the full model, which
will include Yukawa interaction and the self-action of the scalar field.

\section*{Aknowlegement}
Author is grateful to  V A Petrov 
for useful discussion.

\end{document}